\documentclass[conference]{IEEEtran}
\usepackage[dvips]{graphicx}
\usepackage{float}	
\usepackage{cite}	
\usepackage{subfigure}
\usepackage{amssymb}
\usepackage{amsfonts}
\usepackage[center]{caption}
\usepackage{cite}
\usepackage{array}
\usepackage{slashbox,booktabs,amsmath}
\usepackage{xcolor}
\usepackage{amsmath}
\usepackage{amssymb}
\usepackage{times}		
\usepackage{url}
\usepackage{mathrsfs}
\usepackage{mathdots}
\usepackage{epsfig}
\usepackage{epstopdf}
\usepackage{enumerate}
\IEEEoverridecommandlockouts

\begin{document}
\title{Constellation Design for Multi-color Visible Light Communications}


\author{
    \IEEEauthorblockN{Qian Gao\IEEEauthorrefmark{1}, Chen Gong\IEEEauthorrefmark{1}, Rui Wang\IEEEauthorrefmark{2}, Zhengyuan Xu\IEEEauthorrefmark{1}, and Yingbo Hua\IEEEauthorrefmark{3}}
    \IEEEauthorblockA{\IEEEauthorrefmark{1}University of Science and Technology of China, Hefei, Anhui, China
    \\}
    \IEEEauthorblockA{\IEEEauthorrefmark{2}Tongji University, Shanghai, China
    \\}
    \IEEEauthorblockA{\IEEEauthorrefmark{3}University of California at Riverside, California, USA
    \\e-mail: qgao@ustc.edu.cn, cgong821@ustc.edu.cn, liouxingrui@gmail.com, xuzy@ustc.edu.cn, yhua@ee.ucr.edu}}


\maketitle

\IEEEpeerreviewmaketitle

%

\maketitle

\begin{abstract}
In this paper, we propose a novel high dimensional constellation design scheme for visible light communication (VLC) systems employing red/green/blue light emitting diodes (RGB LEDs). It is in fact a generalized color shift keying (CSK) scheme which does not suffer efficiency loss due to a constrained sum intensity for all constellation symbols. Crucial lighting requirements are included as optimization constraints. To control non-linear distortion, the optical peak-to-average-power ratio (PAPR) of LEDs is individually constrained. Fixing the average optical power, our scheme is able to achieve much lower bit-error rate (BER) than conventional schems especially when illumination color is more ``unbalanced''. When cross-talks exist among the multiple optical channels, we apply a singular value decomposition (SVD)-based pre-equalizer and redesign the constellations, and such scheme is shown to outperform post-equalized schemes based on zero-forcing or linear minimum-mean-squared-error (LMMSE) principles. To further reduce system BER, a binary switching algorithm (BSA) is employed the first time for labeling high dimensional constellation. We thus obtains the optimal bits-to-symbols mapping.

\end{abstract}

\IEEEpeerreviewmaketitle

\section{Introduction}
In recent years, indoor visible light communication by light emitting diodes (LEDs) has attracted extensive academic attention \cite{Watson,Chen} (and references therein), driven by advancements in designing and manufacturing of LEDs \cite{Kim}. Adoption of LEDs as lighting source can significantly reduce energy consumption and at the same time offering high speed wireless communication, which is the primary focus of visible light communication (VLC) research \cite{Elgala,Vucic,Wang}. Most of the existing schemes employ blue LEDs with a yellow phosphor coating, while with red/green/blue (RGB) LEDs higher data rate is possible because of wavelength division multiplexing.

With RGB LEDs, color-shift keying (CSK) was recommended by the IEEE 802.15.7 Visible Light Communication Task Group \cite{IEEE11}. A few authors have promoted this idea by designing constellations using signal processing tools. Drost et al. proposed an efficient constellation designed for CSK based on billiard algorithm \cite{Drost10}. Monteiro et al. designed the CSK constellation using an Interior Point Method, operating with peak and color cross-talk constraints \cite{Monteiro}. Bai et al. considered the constellation design for CSK to minimize the bit error rate (BER) subject to some lighting constraints \cite{Bai12}.

Despite the fact that the three-dimensional constellation design problems have been formulated in \cite{Drost10,Monteiro,Bai12},
a few important questions have not been addressed.
They include how to compare a system with CSK employed and a conventional decoupled system, the constellation design,
and the peak-to-average power ratio (PAPR) reduction~\cite{Yu}.
In this paper, we propose a novel constellation design scheme in high dimensional space, termed CSK-Advanced.
In our design, arbitrary number of red, blue, and green LEDs can be selected.
With any average optical intensity and average color selected, we formulate an optimization problem to minimize the system symbol error rate (SER) by maximizing the minimum Euclidean distance (MED) among designed symbol vectors. Further, other important lighting factors such as color rendering index (CRI) and luminous efficacy rate (LER) are also considered.
Further, optical PAPR is included as an additional constraint.

The remainder of this paper is organized as follows.
In Section II, we consider the constellation design problem assuming ideal channel.
In Section III, we consider the constellation design for channel with cross-talks (CwC).
An SVD-based pre-equalizer is applied and the constellations are redesigned subject to a transformed set of constraints.
In Section IV, we discuss the optimization of constellations under arbitrary color illuminations. In Section V, we compare our scheme with a decoupled scheme and provide performance evaluation. Finally, Section VI provides conclusions.

%

\section{Constellation Design with Ideal Channel}
\begin{figure}[h]
\centerline{\includegraphics[width=1.1\columnwidth]{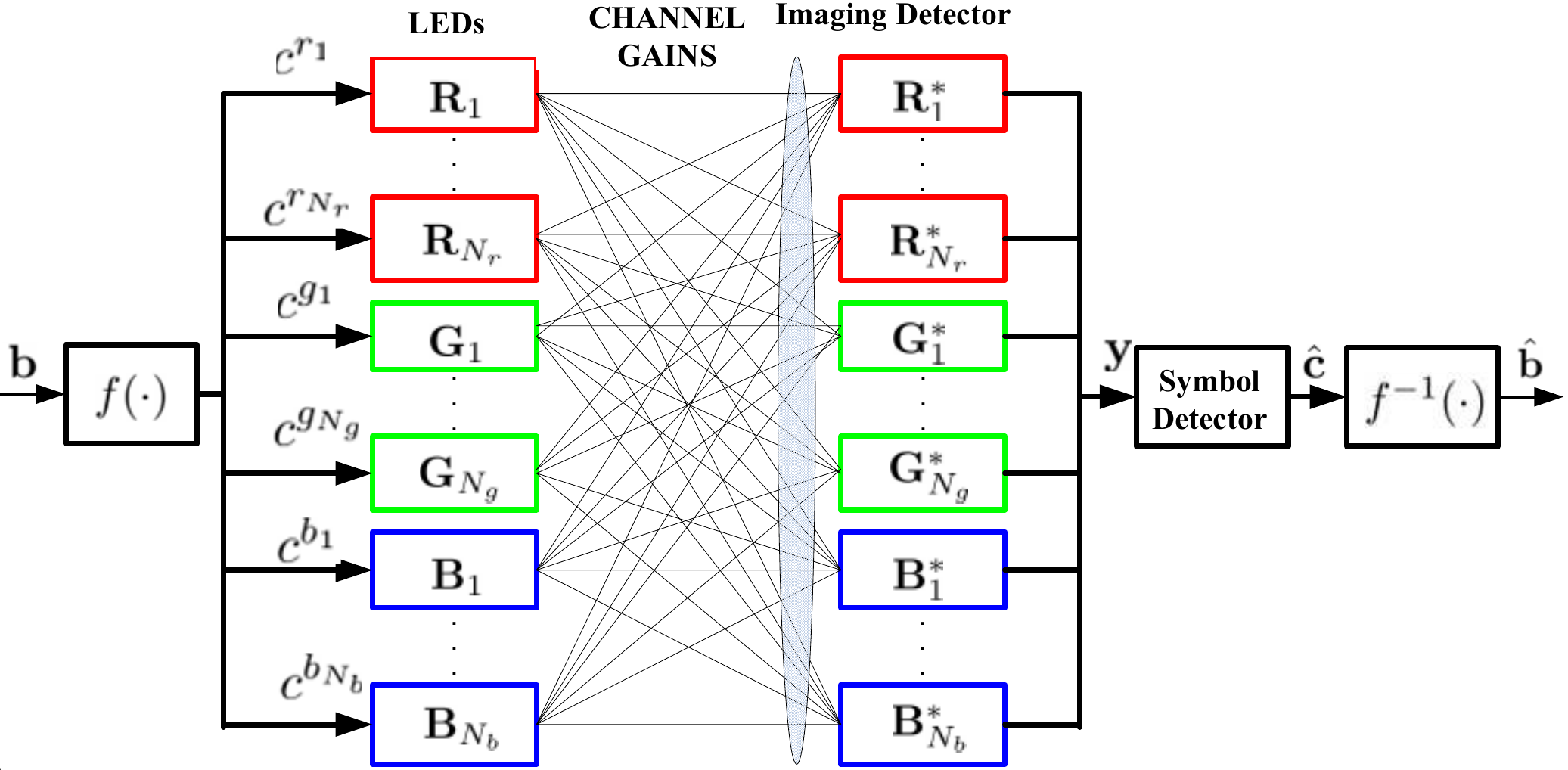}}
\caption{System diagram of the proposed CSK-Advanced System.}
\end{figure}
The system diagram is shown in Fig. 1, with $N_r$ red LEDs, $N_g$ green LEDs, and $N_b$ blue LEDs\footnote{Commercialized products as such are available to our knowledge.}.  In one symbol interval of length $T_s$, a random bit sequence $\mathbf{b}$ of size $N_B\times 1$ is first mapped by a BSA mapper $f(\cdot)$ to a symbol vector $\mathbf{c}$ of size $N_T\times 1$, where $N_T=N_r+N_g+N_b$. The symbol $\mathbf{c}$ is chosen from a constellation
\begin{align}
\mathcal{C}=(\mathbf{c}_1,\mathbf{c}_2,\ldots,\mathbf{c}_{N_c}),
\end{align}
where $N_c=2^{N_B}$ denotes the constellation size. Each component $c_{i,j}$ is applied to the corresponding LED as intensity to transmit, such that $\mathbf{c}_i\geq 0$. The intensity vector $\mathbf{c}_i$ is then multiplied with the optical channel $\mathbf{H}$ of size $N_T\times N_T$.
\footnote{We assume perfect channel knowledge in this paper and do not account for channel estimation errors.}
The output of the color filters can be written as follows,
\begin{equation}
\mathbf{y}=\gamma\eta\mathbf{H}\mathbf{c}+\mathbf{n},
\end{equation}
where $\eta$ is the electro-optical conversion factor, $\gamma$ is the photodetector responsivity.
Without loss of generality (w.l.o.g.), assume $\gamma\eta=1$.
The noise $\mathbf{n}$ is the combination of shot noise and thermal noise~\cite{Zeng09}, assuming
$\mathbf{n} \sim {\cal N}({\bf 0}, {\bf I} \cdot N_0/2)$.
It should be noted that the imaging detector is followed by imperfect color filters such that cross-talks may exist.
The received intensity vector $\mathbf{y}$ is passed through a symbol detector to obtain an estimate of the transmitter symbol,
which is then de-mapped by $f^{-1}(\cdot)$ to recover the bit sequence.
We assume line-of-sight (LOS) links without inter-symbol interference.

We first consider ideal channel, i.e. $\mathbf{H}=\mathbf{I}_{N_T}$. Define a joint constellation vector $\boldsymbol{\mathbf{c_T}}=[\mathbf{c}_1^T~\mathbf{c}_2^T~\ldots\mathbf{c}_{N_c}^T]^T$, and the $i$-th symbol is written as
\begin{align}
\mathbf{c}_i&=[c_i^{r_1}\ldots c_i^{r_{N_r}}~c_i^{g_1}\ldots c_i^{g_{N_g}}~c_i^{b_1}\ldots c_i^{b_{N_b}}]^T=\mathbf{J}_i\boldsymbol{\mathbf{c_T}},
\end{align}
where $\mathbf{J}_i=[\mathbf{O}_{N_T}\ldots \mathbf{I}_{N_T}\ldots \mathbf{O}_{N_T}]$ is a selection matrix with all zeros except for an identity matrix at the $i$-th block.

\subsection{The objective function}
Our objective is to minimize the system SER subject to several visible lighting constraints.
We aim to max the minimum MED $d_{min}$, i.e., maximize $t$ such that the following holds for all $l$~\cite{Beko12}
\begin{equation}
\mathbf{c}^T_T\mathbf{F}_l\boldsymbol{\mathbf{c_T}}\geq t.\label{8}
\end{equation}
where the parameter $t$ will be optimized and we obtain $d_{min}$ through this optimization. $\mathbf{F}_{l(p,q)}=\mathbf{E}_{pq}$, $\mathbf{E}_p=\mathbf{e}_p^T\otimes \mathbf{I}_{N_c}$ (Kronecker product), $\mathbf{e}_p$ of size $N_T\times 1$ has all zeros except the $p$-th element being one, $\mathbf{E}_{pq}=\mathbf{E}_p^T\mathbf{E}_p-\mathbf{E}_p^T\mathbf{E}_q-\mathbf{E}_q^T\mathbf{E}_p+\mathbf{E}_q^T\mathbf{E}_q$, and
\begin{equation}
l\cong(p-1)N_c-\frac{p(p+1)}{2}+q,~p,q\in 1,2,\ldots,N_c,~p<q.\label{26}
\end{equation}
The distance constraints \eqref{8} are nonconvex in $\boldsymbol{\mathbf{c_T}}$. We approximate \eqref{8} by a first order Taylor series approximation around $\boldsymbol{\mathbf{c_T}}^{(0)}$, i.e.
\begin{align}
\mathbf{c}^T_T\mathbf{F}_l\boldsymbol{\mathbf{c_T}}&\cong2\boldsymbol{\mathbf{c_T}}^{(0)T}\mathbf{F}_l\boldsymbol{\mathbf{c_T}}-\boldsymbol{\mathbf{c_T}}^{(0)T}
\mathbf{F}_l\boldsymbol{\mathbf{c_T}}^{(0)}\triangleq h_l^{(0)}(\boldsymbol{\mathbf{c_T}})\geq t, \label{28}
\end{align}
where $\boldsymbol{\mathbf{c_T}}^{(0)}$ is either a random initialization point or a previously attained estimate.

\subsection{The average color and average power constraint}
A designer may wish to constrain the average color,
as non-white illumination could be useful in many places. The average of all LEDs' intensities can be written as the following $N_T\times 1$ vector
\begin{align}
\bar{\mathbf{c}}&=\big(\frac{1}{N_c}\sum_{i=1}^{N_c}\mathbf{J}_i\big)\boldsymbol{\mathbf{c_T}}= \bar{\mathbf{J}}\boldsymbol{\mathbf{c_T}}\notag\\
&=[\bar{c}^{r_1}\ldots\bar{c}^{r_{N_r}}~\bar{c}^{g_1}\ldots\bar{c}^{g_{N_g}}~\bar{c}^{b_1}\ldots\bar{c}^{b_{N_b}}]^T.
\end{align}
We consider the average power of each color, i.e., a $3\times 1$ vector $\mathbf{c}_3$ given as follows,
\begin{equation}
\mathbf{c}_3=P_o\cdot [\bar{c}^r~\bar{c}^g~\bar{c}^b]^T=\mathbf{K}\bar{\mathbf{c}}=\mathbf{K}\bar{\mathbf{J}}\boldsymbol{\mathbf{c_T}},
\end{equation}
where $\mathbf{K}$ is a selection matrix summing up r/g/b intensities accordingly, $P_o$ is the average optical power, and
\begin{equation}
\bar{c}^r+\bar{c}^g+\bar{c}^b=1,
\end{equation}
\begin{equation}
P_o\bar{c}^x=\bar{c}^{x_1}+\ldots+\bar{c}^{x_{N_x}},
\end{equation}
where $x\in \{r,g,b\}$. By properly selecting $\mathbf{c}_3$, the CRI and LER constraints can be met \cite{Broadbent}.

\subsection{The optical PAPR constraint}
For each LED, the optical PAPR is defined as the ratio of the highest power over the average power.
Mathematically, the PAPR of the $j$-th LED can be written as follows,
\begin{equation}
\Phi_j=\frac{\max(\mathbf{K}_j\boldsymbol{\mathbf{c_T}})}{1/N_c \cdot\ \sum(\mathbf{K}_j\boldsymbol{\mathbf{c_T}})},~~\forall j\in[1,N_T],
\end{equation}
where $\sum(\mathbf{a})$ denotes the summation of all elements of vector, $\mathbf{K}_j$ is a selection matrix of size $N_c\times N_cN_T$,  $\max(\mathbf{K}_j\boldsymbol{\mathbf{c_T}})$ denotes the largest element of vector $\mathbf{K}_j\boldsymbol{\mathbf{c_T}}$.

The PAPR of an individual LED can be constrained as follows
\begin{equation}
\Phi_j\leq \alpha_j,~~\forall j\in[1,N_T].
\end{equation}

\subsection{CRI and LER constraints}
CRI stands for a quantitative measure of ability of light sources to reproduce the colors of objects faithfully, comparing with an ideal lighting source \cite{CIE}. LER measures how well light sources creates visible light. It is the ratio of luminous flux to power. Depending on context, the power can be either the radiant flux of the source's output, or it can be the total power (electric power, chemical energy, or others) consumed by the source \cite{Stimson}. The CRI and LER are important practical lighting constraints. By properly selecting $\mathbf{c}_3$, specific CRI and LER constraints can be met.

\subsection{The optimization problem}
When $\mathbf{H}=\mathbf{I}_{N_T}$ the problem can be formulated as follows,
\begin{equation}
\begin{aligned}
& \underset{\boldsymbol{\mathbf{c_T}},t}{\text{max}}
& & t \\
& \text{s.t.}
& & \mathbf{K}\bar{\mathbf{J}}\boldsymbol{\mathbf{c_T}}=\mathbf{c}_3,\\
&&& \boldsymbol{\mathbf{c_T}}\geq 0,\\
&&& h_l^{(0)}(\boldsymbol{\mathbf{c_T}})\geq t \qquad \forall l,\\
&&& N_c\max(\mathbf{K}_j\boldsymbol{\mathbf{c_T}})-\alpha_j\sum(\mathbf{K}_j\boldsymbol{\mathbf{c_T}})\leq 0, \qquad \forall j,
\end{aligned}\label{34}
\end{equation}
which can be straightforwardly proven as a convex optimization problem. With the first three constraints, it is termed as a regular optimization problem and with all constraints a PAPR-constrained problem. By iteratively solving \eqref{34}, a local optimal constellation $\boldsymbol{\mathbf{c_T}}^1$ can be obtained \footnote{One can refer to a similar problem in \cite{Beko12} for convergence, complexity and performance analysis.}. With multiple runs starting from different initial point $\boldsymbol{\mathbf{c_T}}^{(0)}$, the best of solutions, $\boldsymbol{\mathbf{c_T}}^*$ is selected.

\section{Constellation Design with CwC}
The channel cross-talks exist when the transmitting LED's emission spectral does not match the receiver filter's transmission spectral.
It can be described by the following structure assuming single RGB LED is employed based on \cite{Drost10,Monteiro} and experiments,
\begin{align*}
\mathbf{H}_c=
&\begin{bmatrix}
1-\epsilon & \epsilon & 0\\
\epsilon & 1-2\epsilon & \epsilon\\
0 & \epsilon & 1-\epsilon \\
\end{bmatrix},\label{15}
\end{align*}
where the parameter $\epsilon\in[0,0.5)$ characterizes both attenuation and interference effects.

By singular value decomposition (SVD), $\mathbf{H}_c=\mathbf{U}\mathbf{S}\mathbf{V}^H$,
where $\mathbf{U}$ and $\mathbf{V}$ are unitary matrices of size $N_T\times r$, $\mathbf{S}$ is a diagonal matrix of size $rank(\mathbf{H}_c)\times rank(\mathbf{H}_c)$.
In this case, $r$ is the dimension of space for constellation design instead of $N_T$. We apply a pre-equalizer $\mathbf{P}=\mathbf{V}\mathbf{S}^{-1}$ at the transmitter-side and a post-equalizer $\mathbf{U}^H$ at the receiver-side to equalize the channel\footnote{$\mathbf{n}$ and $\mathbf{U}^H\mathbf{n}$ have the same distribution, since $\mathbf{U}^H$ is unitary.}. Define $\boldsymbol{\mathbf{P_T}}=\mathbf{I}_{N_c}\otimes \mathbf{P}$, and the optimization in \eqref{34} can be transformed as
\begin{equation}
\begin{aligned}
& \underset{\boldsymbol{\mathbf{c_T}},t}{\text{max}}
& & t \\
& \text{s.t.}
& & \mathbf{K}\bar{\mathbf{J}}\boldsymbol{\mathbf{P_T}}\boldsymbol{\mathbf{c_T}}=\mathbf{c}_3,\\
&&& \boldsymbol{\mathbf{P_T}}\boldsymbol{\mathbf{c_T}}\geq 0,\\
&&& h_l^{(0)}(\boldsymbol{\mathbf{c_T}})\geq t \qquad \forall l,\\
&&& N_c\max(\mathbf{K}_j\boldsymbol{\mathbf{P_T}}\boldsymbol{\mathbf{c_T}})-\alpha_j\text{sum}(\mathbf{K}_j\boldsymbol{\mathbf{P_T}}\boldsymbol{\mathbf{c_T}})\leq 0,
~ \forall j.
\end{aligned}\label{35}
\end{equation}
It should be noted that $\boldsymbol{\mathbf{c_T}}$ now is of dimension $N_cr\times 1$, i.e., the constellation is designed in a $r$-dimensional space.

To further minimize the system BER with a fixed SER, a good bit-to-symbol mapping function $f(\cdot)$ as shown in Fig. 1 need to be designed.
In this paper, we apply the binary switching (BSA) algorithm to optimize the mapping. Since it is not the main focus of this paper, the details of BSA are omitted (we refer the readers interested to \cite{Schreckenbach}).

\section{Optimized Constellations with Targeted Color Illumination}
We provide numerical illustration of advantages of the CSK-Advanced with one RGB LED, i.e. $N_r=N_g=N_b=1$. Both the CSK-Advanced and the conventional decoupled scheme can work with arbitrary color illumination. With one RGB LED, $c_i~(i\in[1,3])$ for the decoupled scheme takes value from OOK (2-PAM) constellations. To make a fair comparison, an 8-CSK-Advanced constellation is designed,
with equal spectrum efficiency (bits/sec/Hz), equal average optical powers, and equal average color.

\subsection{Constellation design with ideal channel}

\subsubsection{Balanced lighting system}
If the average intensity of each color is similar, we call the corresponding system ``Balanced lighting system''. For example, we choose average color as $\mathbf{c}_3^B=10\cdot[1/3,~1/3,~1/3]^T$ and the average power $P_o=10$. For the conventional scheme, each LED can simply take value independently
from the following binary constellations

\begin{equation*}
\mathcal{C}_{B,r}=\mathcal{C}_{B,g}=\mathcal{C}_{B,b}=[0,6.67].
\end{equation*}

The MED for each branch is $d_{min}\approxeq 6.67$. For our scheme, the optimized constellation is as follows (column 1 to 4 and column 5 to 8 are separated
due to space limit.)

\begin{align*}
\mathcal{C}_B^8(:,1:4)=
\begin{bmatrix}
    0  &  0 &   0  &  4.8485  \\
   14.5455  &  0 &  7.2727  &  4.8485 \\
    0 &  14.5455  &  0  &  4.8485 &
\end{bmatrix},\label{15}
\end{align*}
\vspace{.1in}
\begin{align*}
\mathcal{C}_B^8(:,5:8)=
\begin{bmatrix}
    0  &  0 &  14.5455 &   7.2727\\
    0  &  0  &  0  &  0\\
    0  & 7.2727   & 0  &  0
\end{bmatrix}.
\end{align*}

The MED equals $7.27$, such that we could expect a lower SER with sufficient SNR. The asymptotic power gain is approximately $0.86$dB (=$10\times \log(7.27/6.67)$).
\vspace{.1in}

\subsubsection{Unbalanced lighting system}
We choose the average color as  $\mathbf{c}_3^U=10\cdot[0.44,~0.33,~0.22]^T$. With the conventional scheme, the LEDs take value from constellations

\begin{equation*}
\mathcal{C}_{U,r}=[0,8.88]~~\mathcal{C}_{U,g}=[0,6.66]~~\mathcal{C}_{U,b}=[0,4.44].
\end{equation*}

With our scheme, the optimized constellation is as follows

\begin{align*}
\mathcal{C}^8_u(:,1:4)=
\begin{bmatrix}
    0  &  7.2590 &  14.5550  &  0  \\
    6.4859  &  0  &  0  & 12.9718  \\
    3.2598  &  7.2589  &  0  & 0
\end{bmatrix},
\end{align*}
\vspace{.1in}
\begin{align*}
\mathcal{C}^8_u(:,5:8)=
\begin{bmatrix}
    6.4454  &  0 &  7.2960  &  0 \\
    7.2090  &  0  &  0  & 0 \\
    0  &  7.2590  &  0  & 0
\end{bmatrix}.
\end{align*}

The MED is approximately $7.26$, which is smaller than MED of one branch but larger than MEDs of two branches of the conventional scheme.
\vspace{.1in}

\subsubsection{Extremely Unbalanced lighting system}
We choose the average color as  $\mathbf{c}_3^E=10\cdot[0.7,~0.15,~0.15]^T$. With the conventional scheme, the LEDs take value from constellations

\begin{equation*}
\mathcal{C}_{E,r}=[0,14]~~\mathcal{C}_{E,g}=[0,3]~~\mathcal{C}_{E,b}=[0,3].
\end{equation*}

With our scheme, the optimized constellation is

\begin{align*}
\mathcal{C}^E_8(:,1:4)=
\begin{bmatrix}
 12.6277   & 0  &  0  &  6.3139 \\
    0 &   0  &  6.3139  &  0 \\
    0  &  0 &   0 &  0
\end{bmatrix},
\end{align*}
\vspace{.1in}
\begin{align*}
\mathcal{C}^E_8(:,5:8)=
\begin{bmatrix}
  9.0584  &  0  & 9.0584 &  18.9416\\
  0 &  0 &  5.6861  &  0\\
  5.6861  &  6.3139   & 0  &  0
\end{bmatrix}.
\end{align*}

The MED is approximately $6.31$, which is smaller than MED of one branch but larger than MEDs of two branches of the conventional scheme.

\subsubsection{PAPR-constrained system}
\begin{table}[H]
  \centering
  \caption{MED with varying PAPR and average color.}
  \vspace{-0.2in}
  \begin{tabular}{ccccc} \\ \hline
    $d_{min}$ & $\mathcal{C}_B^8$ & $\mathcal{C}_U^8$ & $\mathcal{C}_E^8$ \\ \hline
    $\alpha=1.5$ & 3.54 & 3.40 & 2.84  \\
    $\alpha=2$ & 6.67 & 5.58 & 4.38  \\
    $\alpha=4$ & 7.07 & 7.26 & 6.31 \\
    $\alpha=6$ & 7.27 & 7.26 & 6.31  \\ \hline
  \end{tabular}
\end{table}

If \textit{identical individual PAPR constraints} are added, i.e. $\alpha_j=\alpha$ into the optimization. The corresponding MEDs with varying PAPR are summarized in Table I.
It can be observed that there is a tradeoff between minimum distance and PAPR for the three cases. With extremely low PAPR, e.g. $\alpha=1.5$, the system suffers from severe power loss. With a PAPR increase of 3dB, e.g. from $\alpha=2$ to $\alpha=4$ which is typically tolerable, the power gain of unbalance systems are larger than balanced system.


\section{Performance Evaluation}
We simulate using bit sequence of length $N=9 \times 10^6$ for selected cases above to compare the BER performance among different systems versus different optical SNR, defined as \cite[Eq.27]{Karout1} as follows,
\begin{equation}
\gamma_o=10\log_{10}\frac{P_o}{\sqrt{N_bN_0}}.
\end{equation}

Selected BER curves versus optical SNR are included in Fig. 2 and Fig. 3. In Fig. 2, CSK-Advanced system applies constellation $\mathcal{C}_8(\mathbf{c}_3^B)$ and in Fig. 3 constellation $\mathcal{C}_8(\mathbf{c}_3^E)$ is used. It can be observed that with CSK-Advanced scheme non-trivial power gain is obtained over the conventional system, especially when the average color is not balanced. The optimized mapping by BSA offers additional power gain for all OSNR range.

\begin{figure}[H]
\centerline{\includegraphics[width=1.1\columnwidth]{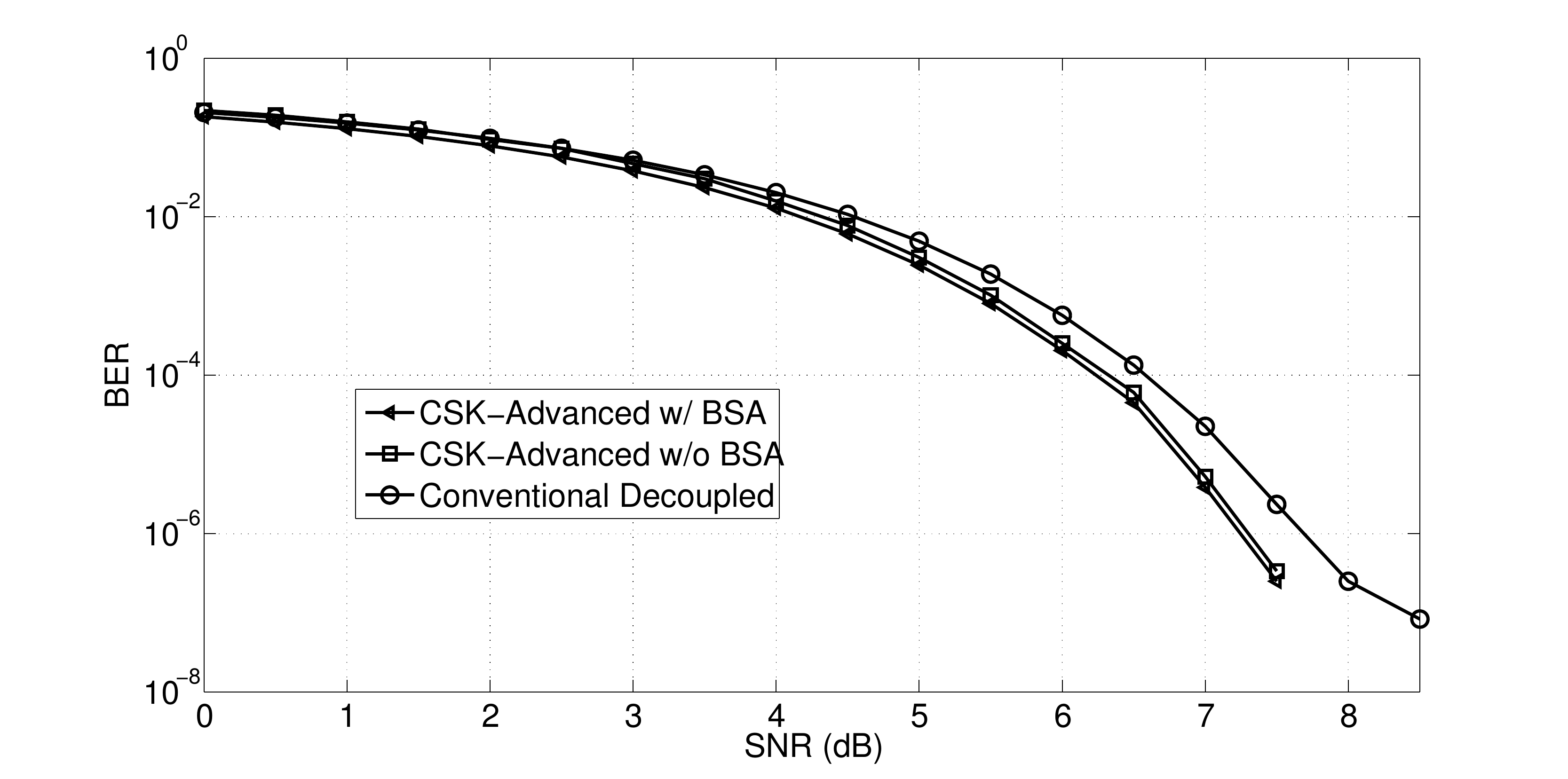}}
\caption{Balanced conventional system vs CSK-Advanced systems.}
\end{figure}
\vspace{-0.2in}
\begin{figure}[H]
\centerline{\includegraphics[width=1.1\columnwidth]{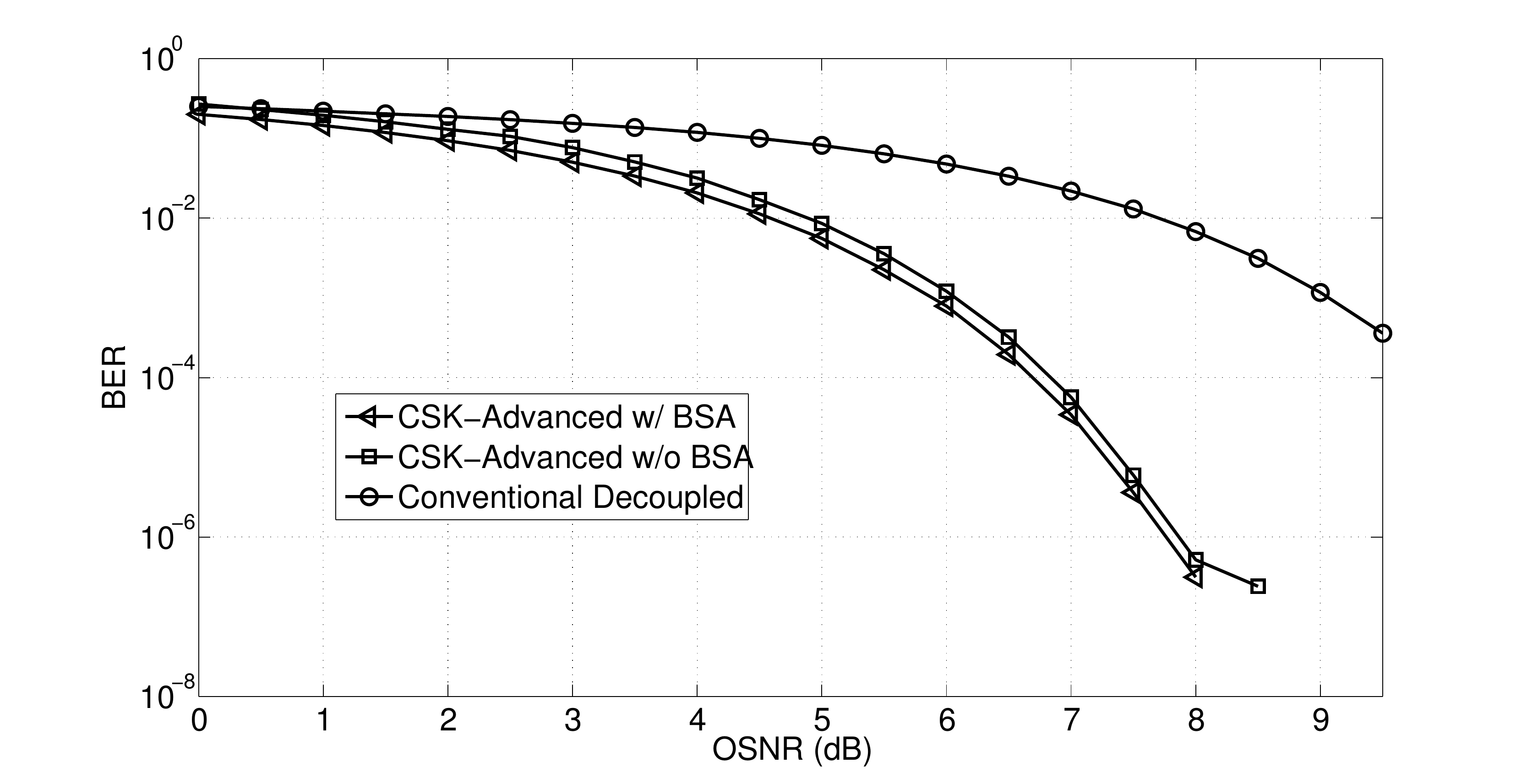}}
\caption{Extremely Unbalanced conventional system vs CSK-Advanced systems.}
\end{figure}

Histogram of MEDs of 1000 local optimal constellations for the balanced system is shown by Fig. 4. It can be seen that approximately $1/4$ of the runs will converge to satisfactory MEDs. Therefore, we would suggest only 20$-$30 runs in practice to reduce complexity.
\begin{figure}[H]
\centerline{\includegraphics[width=1.1\columnwidth]{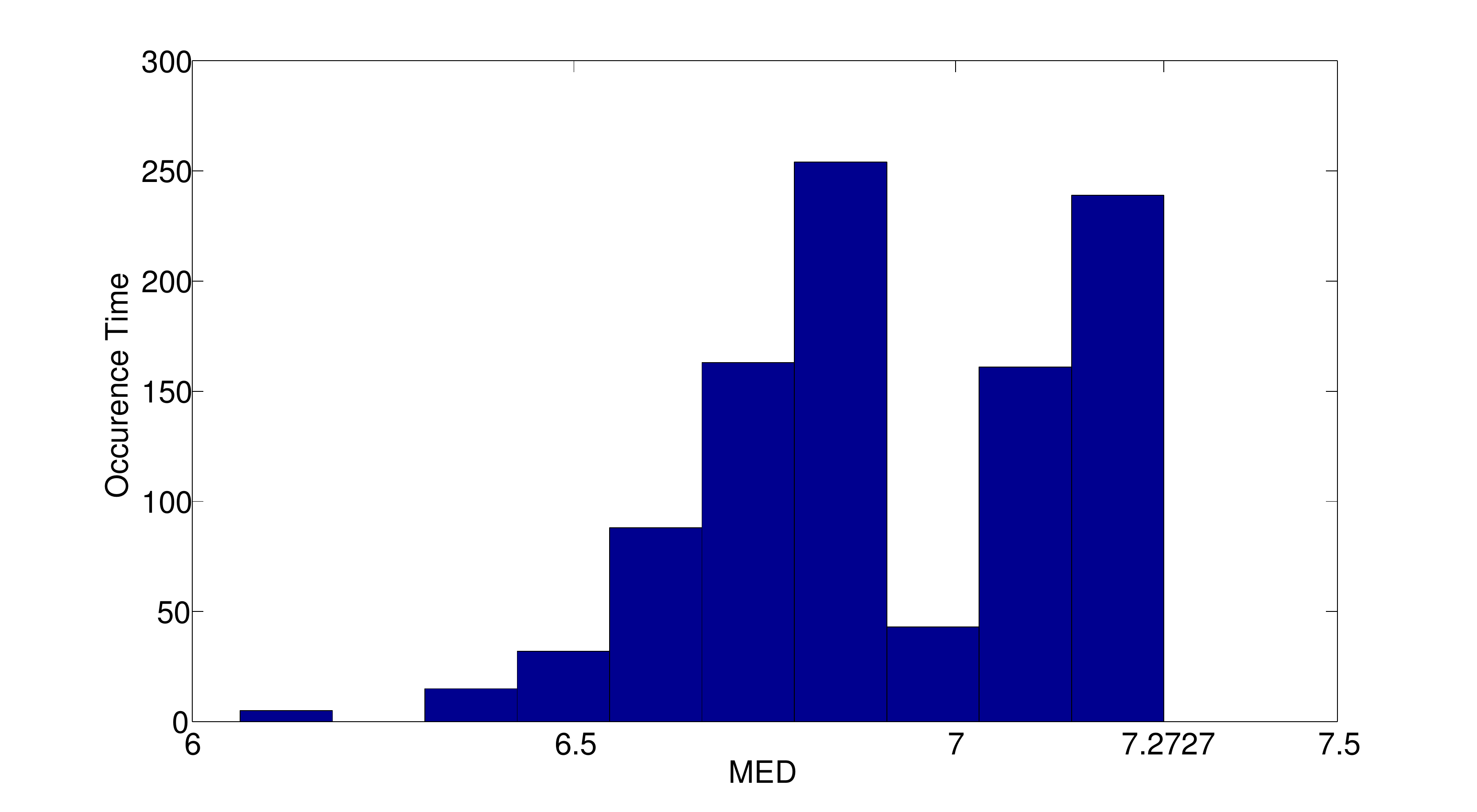}}
\caption{Histogram of MEDs of 1000 local optimal constellations.}
\end{figure}

\subsection{Constellation labeling}

The optimized bit-to-symbol mapping obtained by the BSA when $\gamma_o=5$dB is included in Table II. With optimized mapping, only 1.33 out of 3 bits on average are in error when a symbol error occurs. Without BSA based mapping how average, 1.73 out of 3 bits on average (over results observed from 100 random labelings) are mis-interpreted instead. The optimized mapping tables can be computed offline.

\begin{table}[h]
  \centering
  \caption{Optimized Bit-to-symbol mapping with OSNR=5dB.}
   \vspace{-0.2in}
  \begin{tabular}{ccccccccc} \\ \hline
    Constellation Point & Optimized Labeling \\ \hline
    (0,~0,~7.27) & 000 \\
    (0,~0,~0) & 001 \\
    (0,~14.55,~0) & 010 \\
    (0,~7.27,~0) & 011 \\
    (0,~0,~14.55) & 100 \\
    (7.27,~0,~0) & 101 \\
    (14.55,~0,~0) & 110 \\
    (4.85,~4.85,~4.85) & 111 \\ \hline
  \end{tabular}
\end{table}

\subsection{Constellation design with CwC}
\subsubsection{SVD-based pre-equalizer}
Consider the following $3\times 3$ channel with moderate cross-talks, e.g. $\epsilon_1=0.1$,
\begin{align*}
\mathbf{H}_{\epsilon_1}\notag=
&\begin{bmatrix}
0.9 & 0.1 & 0\\
0.1 & 0.8 & 0.1\\
0 & 0.1 & 0.9\\
\end{bmatrix}.\label{15}
\end{align*}

By SVD we have $\mathbf{H}_{\epsilon_1}=\mathbf{U}_{\epsilon_1}\mathbf{S}_{\epsilon_1}\mathbf{V}_{\epsilon_1}^H$. The pre-equalizer $\mathbf{P}_{\epsilon_1,pr}=\mathbf{V}_{\epsilon_1}\mathbf{S}_{\epsilon_1}^{-1}$ and post-equalizer $\mathbf{P}_{\epsilon_1,po}=\mathbf{U}^H_{\epsilon_1}$. The corresponding optimized constellation for balanced system is
\begin{align*}
\mathcal{C}_{\epsilon_1}^8(:,1:4)=\begin{bmatrix}
   -7.8376  & -5.7208 &  -4.6893 &  0 \\
    8.6391 &  -0.9559 &  -5.1688 &  0 \\
    3.8794  & -2.0338 &   2.3211 &  0
\end{bmatrix},
\end{align*}
\begin{align*}
\mathcal{C}_{\epsilon_1}^8(:,5:8)=\begin{bmatrix}
     -8.4386  & -3.9188 &  -7.8123 &  -7.7708\\
     4.5539 &   4.3196 &   0.0000 &  -0.0742\\
     -2.2190 &   1.9397 &  -7.7338  &  3.8463
\end{bmatrix}.
\end{align*}
The MED with varying area of overlap for balanced, unbalanced, and extremely unbalanced systems are summarized in Table III.
In practice, the mismatch between the emission spectra of the transmitter LEDs and the transmission spectra of the receiver filters is restricted, and cases with $\epsilon\geq 0.2$ are very rare.

\begin{table}[h]
  \centering
  \caption{MED with varying area of overlap.}
  \vspace{-0.2in}
  \begin{tabular}{cccccccc} \\ \hline
    $d_{min}$ & $\mathcal{C}_{B,D}^8$ & $\mathcal{C}_{U,D}^8$ & $\mathcal{C}_{E,D}^8$ \\ \hline
    $\epsilon=0$ & 7.2727 & 7.2590 & 6.3139 \\
    $\epsilon=0.05$ & 6.7621 & 6.6748 & 5.9275  \\
    $\epsilon=0.1$ & 6.3275 & 6.1464 & 5.5657 \\
    $\epsilon=0.15$ & 5.9462 & 5.7769 & 5.1635  \\
    $\epsilon=0.2$ & 5.5670 &  5.3692 & 4.7727  \\ \hline
  \end{tabular}
\end{table}

\subsubsection{Comparison with post-equalized systems}
Instead of redesign the constellations subject to a transformed set of constraints due to employment of a pre-equalizer $\mathbf{P}$, zero-forcing (ZF) $\mathbf{G}_Z$ or linear minimum-mean-squared-error (LMMSE) based post-equalizer $\mathbf{G}_{L}$ can be employed at the receiver \cite{Crauss} to mitigate the cross-talks.

Fig. 5 shows the corresponding BERs against increased crosstalks for a balanced system employing different schemes when OSNR is fixed to 5dB. It is seen that our SVD-based scheme significantly outperforms systems employing either ZF or LMMSE post-equalizers. Fig. 6 shows the BERs against OSNR for a balanced system when $\epsilon$ is fixed to 0.1. With this particular parameters chosen, there is no significant difference between ZF and LMMSE based system performance and therefore we only included the LMMSE based results.

\begin{figure}[H]
\centerline{\includegraphics[width=1.1\columnwidth]{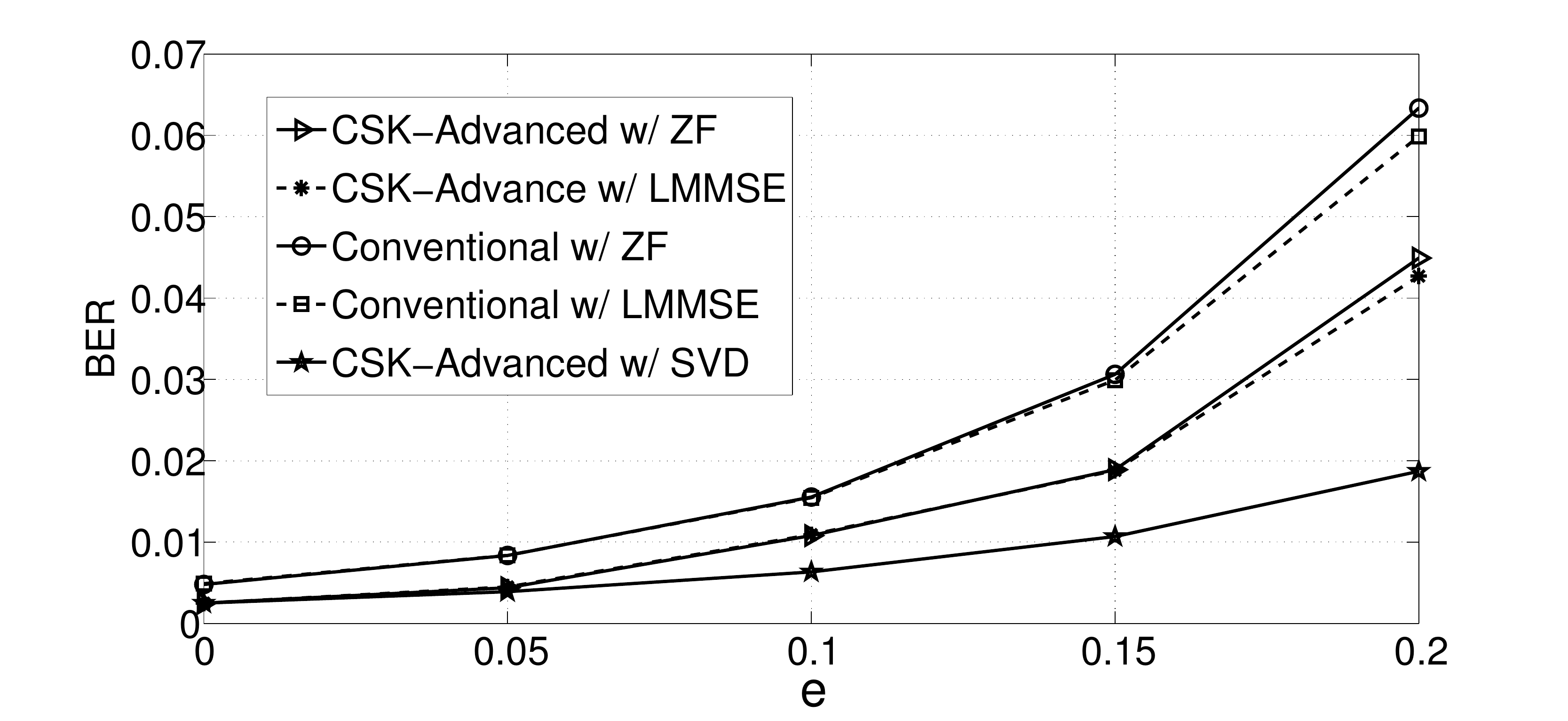}}
\caption{BER against $\epsilon$ with OSNR=5dB for a balanced system.}
\end{figure}
\vspace{-0.2in}
\begin{figure}[H]
\centerline{\includegraphics[width=1.1\columnwidth]{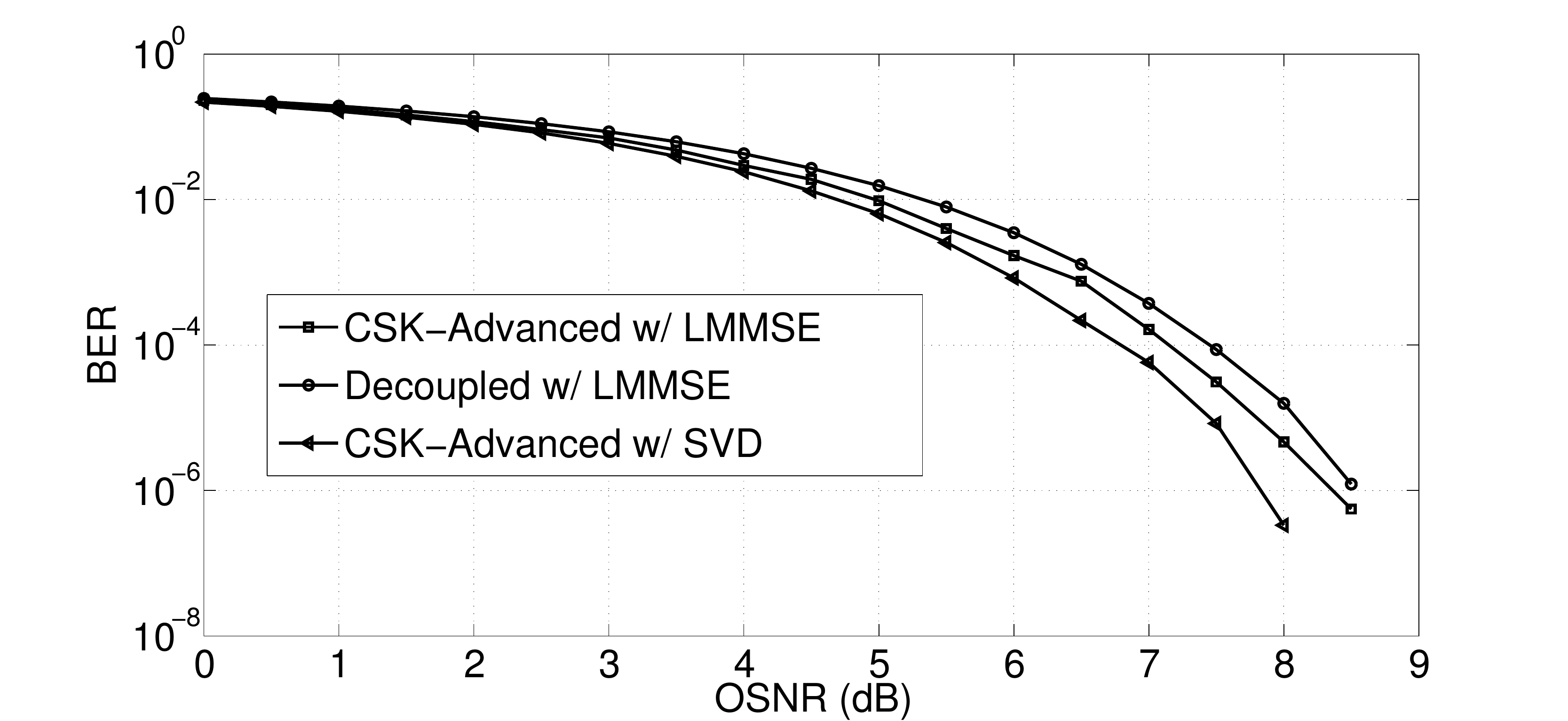}}
\caption{BER against OSNR  with $\epsilon=0.1$ for a balanced system.}
\end{figure}

\section{Conclusion}\label{seccon}
A novel constellation design scheme, named CSK-Advanced, for VLC with arbitrary number of RGB LEDs, is proposed in this paper. With both optimized constellation and bits-to-symbols mapping, significant power gains are observed compared with conventional decoupled systems. For more unbalanced color illumination, the larger power gains can be expected. To avoid excessive nonlinear distortion, optical PAPR constraints is included into the optimization. Furthermore, to deal with CwC, an SVD-based pre-equalizer is introduced. It is shown by simulations that the proposed scheme significantly outperforms various benchmarks
employing ZF or LMMSE-based post-equalizers.


\begin{thebibliography}{9}
\bibitem{Watson} S. Watson, M. Tan, S.P. Najda, P. Perlin, M. Leszczynski, G. Targowski, S. Grzanka, and A. E. Kelly, ``Visible light communications using a directly modulated 422 nm GaN laser diode,'' {\it Opt. Lett.}, vol. 38, no. 19, pp. 3792-3794, 2013.
\bibitem{Chen} C. Chen, P. Wu, H Lu, Y. Lin, J. Wen, and F. Hu, ``
Bidirectional phase-modulated hybrid cable television/radio-over-fiber lightwave transport systems,'' {\it Opt. Lett.}, vol. 38, no. 4, pp. 404-406, 2013.
\bibitem{Kim} J. K. Kim and E. F. Schubert, ``Transcending the replacement paradigm of solid-state lighting,'' {\it Opt. Express}, vol. 16, no. 26, pp. 21835-21842, 2008.
\bibitem{Elgala} H. Elgala and T.D.C. Little, ``
Reverse polarity optical-OFDM (RPO-OFDM): dimming compatible OFDM for gigabit VLC links,'' {\it Opt. Express}, vol. 21, no. 20, pp. 24288-24299, 2013.
\bibitem{Vucic} J. Vucic and K.D. Langer, ``High-speed visible light communications: State-of-the-art,'' {\it OFC/NFOEC}, pp. 1-3, Mar. 2012.
\bibitem{Wang} Y. Wang, Y. Wang, N. Chi, J. Yu, and H. Shang, ``Demonstration of 575-Mb/s downlink and 225-Mb/s uplink bi-directional SCM-WDM visible light communication using RGB LED and phosphor-based LED,'' {\it Opt. Express}, vol. 21, no. 1, pp. 1203-1208, 2013.
\bibitem{IEEE11} IEEE 802.15.7 Visible Light Communication Task Group, https://mentor.ieee.org/802.15/documents?is group=0007.
\bibitem{Drost10} R.J. Drost and B.M. Sadler, ``Constellation design for color-shift keying using billiards algorithms,'' {\it IEEE Globecom Workshop}, pp. 980-984, Dec. 2010.
\bibitem{Monteiro} E. Monteiro and S. Hranilovic, ``Constellation design for color-shift keying using interior point methods,'' {\it IEEE OWC-WS}, pp. 1224-1228, Dec. 2012.
\bibitem{Bai12} B. Bai, Q. He, Z. Xu, and Y. Fan, ``The color shift key modulation with non-uniform signaling for visible light communication,'' {\it IEEE ICCC-WS-OWCC}, pp. 37-42, Aug. 2012.
\bibitem{Yu} Z. Yu, R.J. Baxley, and G.T. Zhou, ``Peak-to-average power ratio and illumination-to-communication efficiency considerations in visible light OFDM systems,'' {\it IEEE ICASSP}, pp. 5397-5401, May. 2013.
\bibitem{Zeng09} L. Zeng, D. O'Brien, H. Minh, G. Faulkner, K. Lee, D. Jung, Y. Oh, and E. Won, ``High data rate multiple input multiple output (MIMO) optical wireless communications using white LED lighting,'' JSAC, vol. 27, no.9, pp. 1654-1662, Dec. 2009.
\bibitem{Broadbent} A.D. Broadbent, ``A critical review of the development of the CIE1931 RGB color-matching functions,'' {\it Color Research and Applications}, vol. 29, no. 4, pp. 267-272, Aug. 2004.
\bibitem{Zeger} K. Zeger and A. Gersho, ``Pseudo-gray coding,'' {\it IEEE Trans. Commun.}, vol. 38, no. 12, pp. 2147-2158, Dec. 1990.
\bibitem{Beko12} M. Beko and R. Dinis, ``Designing good multi-dimensional constellations,'' {\it IEEE Wireless Commun. Lett.}, vol. 1, no. 3, pp. 221-224, 2012.
\bibitem{Schreckenbach} F. Schreckenbach, N. Gortz, J. Hagenauer, and G. Bauch, ``Optimization of symbol mappings for bit-interleaved coded modulation with iterative decoding,'' {\it IEEE Commun. Lett.}, vol. 7, no. 12, pp. 593-595, Dec. 2003.
\bibitem{Karout1} J. Karout, E. Agrell, K. Szczerba, and M. Karlsson, ``Optimizing constellations for single-subcarrier intensity-modulated optical systems,'' {\it IEEE Trans. Inf. Theory}, vol. 58, no. 7, pp. 4645-4659, Apr. 2012.
\bibitem{CIE} CIE (1999), ``Colour rendering (TC 1-33 closing remarks),'' {\it Publication 135/2, Vienna: CIE Central Bureau, ISBN 3-900734-97-6}.
\bibitem{Stimson} A.~Stimson, ``Photometry and radiometry for engineers,'' {\it New York: Wiley and Son}.
\bibitem{Crauss} T.P. Crauss, M.D. Zoltowski, and G. Leus, ``Simple MMSE equalizers for CDMA downlink to restore chip sequence: comparison to zero-forcing and RAKE,'' {\it IEEE ICASSP}, vol. 5, pp. 2865-2868 2000.













\end{thebibliography}
\end{document}